\documentclass{emulateapj}
\usepackage{rotating}

\shorttitle{GTC/OSIRIS spectrum of a faint L subdwarf in UKIDSS}
\shortauthors{Lodieu et al.}

\begin{document}

\title{GTC Osiris spectroscopic identification of a faint L subdwarf in the
UKIRT Infrared Deep Sky Survey}

\author{N. Lodieu\altaffilmark{1,2}, M. R. Zapatero Osorio\altaffilmark{3},
E. L. Mart\'{\i}n\altaffilmark{3}, E. Solano\altaffilmark{3},
M. Aberasturi\altaffilmark{3}}
\altaffiltext{1}{Instituto de Astrof\'\i sica de Canarias, C/ V\'\i a L\'actea
s/n, E-38205 La Laguna, Tenerife, Spain }
\altaffiltext{2}{Departamento de Astrof\'isica, Universidad de La Laguna, E-38205 La Laguna, Tenerife, Spain}
\altaffiltext{3}{Centro de Astrobiolog\'ia (CSIC/INTA), 28850 Torrej\'on de Ardoz, Madrid, Spain}

\email{nlodieu@iac.es}

\begin{abstract}
We present the discovery of an L subdwarf in 234 deg$^{2}$ common
to the UK Infrared Telescope (UKIRT) Infrared Deep Sky Survey Large Area 
Survey Data Release 2 and the Sloan Digital Sky Survey Data Release 3\@.
This is the fifth L subdwarf announced to date, the first one identified
in the UKIRT Infrared Deep Sky Survey, and the faintest known. The blue 
optical and near-infrared colors of ULAS\,J135058.86$+$081506.8 and its 
overall spectra energy distribution are similar to the known mid-L subdwarfs.
Low-resolution optical (700--1000 nm) spectroscopy with the Optical
System for Imaging and low Resolution Integrated Spectroscopy 
spectrograph on the 10.4 m Gran Telescopio de Canarias reveals that 
ULAS\,J135058.86$+$081506.8 exhibits
a strong K\,{\small{I}} pressure-broadened line at 770 nm and a red
slope longward of 800 nm, features characteristics of L-type dwarfs.
From direct comparison with the four known L subdwarfs, we estimate
its spectral type to be sdL4--sdL6 and derive a distance in the
interval 94--170 pc. We provide a rough estimate of the space density
for mid-L subdwarfs of 1.5$\times$10$^{-4}$ pc$^{-3}$.
\end{abstract}


%
%
\section{Introduction}
\label{sdL:intro}

Objects cooler than M dwarfs have been classified into two spectral classes:
L and T \citep{martin99a,kirkpatrick99,geballe02,burgasser06a}. Hundreds of
them are now known as a result of searches of L and T dwarfs in large scale 
surveys\footnote{See DwarfArchives.org for the full list of L and T 
dwarfs}. However, 
at lower metallicity, only four L subdwarfs have been announced to date:
2MASS\,J053253.46$+$824646.5 \citep[2MASS0532; sdL7;][]{burgasser03b},
2MASS\,J162620.34$+$392519.0 \citep[2MASS1626; sdL4;][]{burgasser04},
2MASS\,061640.06$-$640719.4 \citep[2MASS0616; sdL5;][]{cushing09}, and
SDSS\,J125637.13$-$022452.4 \citep[SDSS1256; sdL3.5;][]{sivarani09,burgasser09a}.
The first three were discovered serendipitously in the Two Micron All Sky
Survey \citep[2MASS;][]{cutri03,skrutskie06} archive while the last was
found during a search for ultracool subdwarfs in the Second Data Release (DR)
of the Sloan Digital Sky Survey \citep[SDSS;][]{york00,abazajian05}
spectroscopic database. Increasing the number of ultracool subdwarfs is
essential to study the chemistry in cool atmospheres with low metal content,
the role of metallicity in the shape of the initial mass function
\citep{salpeter55,miller79,scalo86}, and the impact of chemical
composition on the properties of binary stars \citep{riaz08a,jao08,lodieu09b}.

The UKIRT Deep Infrared Sky Survey \citep[UKIDSS;][]{lawrence07}
uses the Wide Field Camera \citep[WFCAM;][]{casali07}
installed on the UK InfraRed Telescope (UKIRT) and the Mauna Kea
Observatory \citep{tokunaga02} photometric system described in
\citet{hewett06}. The pipeline processing is described in
Irwin et al.\ (in preparation)\footnote{Extensive details on the data
reduction of the UKIDSS images is available at
http://casu.ast.cam.ac.uk/surveys-projects/wfcam/technical}
and the WFCAM Science Archive in \citet{hambly08}.
The Large Area Survey (hereafter LAS), one of the UKIDSS components,
will image $\sim$3800 deg$^{2}$ in four filters ($YJHK$) down to
$J$ = 19.6 mag with a significant overlap with SDSS, hence providing
spectral energy distributions from 0.3 to 2.5$\mu$m  for millions of sources.

In this letter, we present the discovery of the first L subdwarf identified
in 234 deg$^{2}$ common to the UKIDSS LAS DR2 \citep{warren07b} and the
SDSS DR3 \citep{abazajian05}. In Section \ref{sdL:sample}, we describe the
photometric and proper motion selections to identify ultracool subdwarfs
in UKIDSS and SDSS\@. In Section \ref{sdL:spectro}, we present the
spectroscopic follow-up observations conducted with the Optical System
for Imaging and low Resolution Integrated Spectroscopy (OSIRIS) spectrograph
mounted on the new 10.4 m Gran Telescopio de Canarias (GTC) telescope
in La Palma (Canary Islands). 
In Section \ref{sdL:analysis}, we discuss the photometric and spectral 
characteristics of the new object, infer its distance, and compute its 
tangential velocity. Finally, we conclude with a discussion on the expected 
number of subdwarfs that could be identified at the completion of UKIDSS\@.

%
%
\section{Sample selection}
\label{sdL:sample}

We have initiated a photometric and proper motion search for ultracool
(spectral types later than M7) subdwarfs by cross-correlating the UKIDSS
and SDSS databases. In the LAS, we have requested only point sources
({\tt{mergedClass}} = $-$1) and good quality detections
({\tt{ppErrBits}} $\leq$ 256). We have employed the following photometric
criteria: $Y-J$ = 0.3--0.75 mag, $J-K \leq$ 0.7 mag because of the strong 
collision-induced H$_{2}$ absorption beyond 2$\mu$m \citep{linsky69}, 
and $z-J \leq$ 2.5 mag because the optical-to-infrared colors of subdwarfs 
are expected to be bluer than their solar metallicity counterparts 
(Section \ref{sdL:analysis}; Figure \ref{fig_sdL:ccd}). We have also
imposed constraints on the LAS detections, requesting $J$ = 14--18.5 mag and
photometric errors less than 0.2 mag in $Y$ and $J$ and better than 0.25 mag
in $H$ and $K$. The bright limit in $J$ was set to place our candidates 
well above the saturation of the UKIDSS LAS\@. In addition, we have 
requested SDSS detections below the 5$\sigma$ limits in $u$ ($\geq$22.3 mag) 
and $g$ ($\geq$ 23.3 mag). Besides the photometric criteria, we have 
imposed a lower limit of 0.18 arcsec yr$^{-1}$ on the proper motion 
\citep{luyten80}, which is measured as the difference between UKIDSS 
and SDSS coordinates.
%
%

This query returned seven photometric candidates. The photometry of
six candidates suggest that they are late-M subdwarfs. Spectroscopic 
follow-up of these six sources is currently underway. However, one object,
ULAS J135058.86$+$081506.8 (hereafter ULAS1350), is the only source with a
negative $J-H$ color and among the faintest ones displaying optical and
near-infrared colors similar to those of 2MASS1626 
\citep[sdL4;][]{burgasser04}, the second L subdwarf known at the time of 
our discovery.
Table \ref{tab_sdL:tab_phot} provides the catalogued photometry of ULAS1350 
and two other known L subdwarfs as well as a field L5 dwarf for comparison. 
Therefore, we prioritized this candidate for optical spectroscopy
that we finally secured with the GTC in 2009\@. From the SDSS/UKIDSS
cross-match, we inferred a total proper motion of 0.280$\pm$0.025 arcsec yr$^{-1}$
thanks to the $\sim$3 yr baseline between both observations (April 2003 and
July 2006). The typical uncertainty on the difference between the
UKIDSS and SDSS astrometry is 0.025 arcsec yr$^{-1}$ for sources brighter than
$J$ = 18.5 mag (see analysis in \citet{dye06} and \citet{lodieu09a}).
This proper motion is lower than the motion of the other four L subdwarfs 
currently known. ULAS1350 is also fainter, suggesting that it lies at a 
larger distance. Its reduced proper motion ($H_{i}$ = 23.45 mag) is 
similar to 2MASS1626 ($H_{i}$ = 23.34 mag) and SDSS1256
($H_{i}$ = 23.41 mag), supporting its possible low-metallicity status.

%
%
\section{Spectroscopic observations}
\label{sdL:spectro}

The 10.4 m GTC started operations in 2009 March at the Observatory del 
Roque de Los Muchachos (La Palma, Canary Islands). It is currently equipped 
with one of its Day One instruments, the OSIRIS instrument \citep{cepa00}. 
The OSIRIS spectrograph consists of two 2048$\times$4096 Marconi CCD42-82 
with an 8 arcsec gap between them and operates
at optical wavelengths, from 365 to 1000 nm. The unvignetted instrument
field of view is 7 $\times$ 7 arcmin with a pixel scale is 0.125 arcsec.
This Letter reports on one of the first scientific results obtained with
GTC/OSIRIS after the gamma-ray burst circular of \citet{castro-tirado09} 
and the observations of SGR 0418$+$5729 by \citet{mignani09}.

We have carried out low-resolution ($R$ $\sim$ 515 at 8500\AA{}) spectroscopy 
of ULAS1350 with the R500R grism available on GTC/OSIRIS and a slit width of 
1 arcsec, projecting onto a full-width-half-maximum of seven pixels onto
the detector, yielding a 480--1000 nm wavelength coverage and a nominal
dispersion of 0.244 nm pixel$^{-}$. The observations were obtained in service 
mode by the GTC staff on 2009 April 30\@. Spectra were taken at parallactic 
angle. Weather conditions were spectroscopic
and seeing around 1.0 arcsec. The total exposure time was 2100 s in one
single exposure. Skyflats and bias frames were observed on 2009 April 4\@. 

All data were reduced under the IRAF\footnote{IRAF is distributed by National 
Optical Astronomy Observatory, which is operated by the Association of 
Universities for Research in Astronomy, Inc., under contract with the 
National Science Foundation.} environment, which includes bias subtraction, 
and flat-field correction using sky-flat images. The individual spectrum 
was optimally extracted and the Earth sky emission lines were removed with 
the APALL task by choosing the appropriate aperture width and background 
intervals. Wavelength calibration (in the air system) was performed with 
an accuracy of 0.05 nm using the internal arc lamp lines of Argon and 
Neon, which were acquired at the end of the night in which ULAS1350 was 
observed. To correct for the instrumental response we used twilight sky-flat 
images (taken with the same instrumental configuration as our target) 
that recorded the solar spectrum. The closest observation of a 
spectrophotometric standard with the R500R grating was carried out on
2009 April 4 \citep[HZ\,44;][]{oke90}, the same night as the skyflats.
We have used it to confirm that 
the correction from instrumental response using the twilight flat images 
is reliable in the wavelength range 700--900 nm, where this star has 
spectrophotometric information in the IRAF database.
Fringing is negligible ($\le$ 1\%) shortwards of 900 nm. The
extracted spectrum of ULAS1350 has a signal-to-noise ratio of around 3
at around 920--925 nm. The usable optical (700--1000 nm) spectrum of 
ULAS1350 smoothed with a boxcar of 51 pixels and a signal-to-noise ratio
of $\sim$20 is displayed in Figure \ref{fig_sdL:sdL_spectrum} and compared 
to the spectra (resolution degraded to that of ULAS1350 and similarly 
smoothed) of the four known L subdwarfs from the literature. 

%
%
\section{Analysis}
\label{sdL:analysis}
\subsection{Photometry}
\label{sdL:analysis_photometry}

From the optical and near-infrared photometry alone, ULAS1350 is unlikely
to be a solar-metallicity object because it shows discrepant colors compared 
to late-M and L dwarfs (Table \ref{tab_sdL:tab_phot}). Its $J-K$ color of 
$-$0.02$\pm$0.16 mag is typical of T dwarfs
and much bluer than solar-metallicity field L dwarfs discovered by 2MASS
and SDSS \citep[$J-K \geq$ 1.2 mag;][]{hawley02}. Moreover, the $z-J$ and 
$i-z$ are also much bluer with values of 1.55$\pm$0.07 and 1.74$\pm$0.11 mag 
compared to $\sim$2.5--3.3 mag and $>$2.0 mag, for mid-L dwarfs with solar 
composition, respectively.

Figure \ref{fig_sdL:sdL_SED} shows the optical to near-infrared spectral 
energy distribution (SED) of ULAS1350 compared to 2MASS1626 
\citep[sdL4;][]{burgasser04} and a solar-metallicity L5 dwarf,
SDSS J144600.60$+$002452.0 \citep{geballe02}. The SED of ULAS1350 clearly
suggests a sub-solar metallicity especially in the near-infrared, as also 
inferred from Figure \ref{fig_sdL:ccd} and Table \ref{tab_sdL:tab_phot}.
To convert observed magnitudes into physical fluxes we have used the zero 
point fluxes corresponding to each SDSS and UKIDSS passband defined in 
\citet{hewett06}. We have flux calibrated the GTC data by integrating the 
observed spectrum convolved with the response curves of the $i$ and $z$ 
filters \citep{fukugita96}. 

\subsection{Spectral features}
\label{sdL:analysis_features}

The GTC spectrum confirms the cool and low-luminosity atmosphere of our 
candidate, for which we determined sdL5$\pm$1 after comparison to the 
four known bright L subdwarfs (Fig. \ref{fig_sdL:sdL_spectrum}).
Indeed, the spectrum of ULAS1350 clearly exhibits a strong pressure-broadened
K\,{\small{I}} band at $\sim$770 nm (Figure \ref{fig_sdL:sdL_spectrum})
as well as a red slope longwards of 800 nm, features typical of early to
mid-L field dwarfs. We also detect the hydride bands of CrH ($\sim$860 nm)
and FeH ($\sim$870 nm and $\sim$990 nm), typically stronger features 
in the spectra of lower metallicity stars \citep{mould76}.
From Figure \ref{fig_sdL:sdL_spectrum}, our spectrum shows a 
decreasing flux at wavelengths 
below 730 nm, a feature that is also shared by the known sdL3.5--sdL5 
objects. It is likely due to absorption by TiO gas, which is usually 
stronger in the subdwarfs relative to solar L dwarfs (probably an effect 
due to inhibited dust formation in low-metallicity atmospheres
\citep[e.g.][]{burgasser03b,reiners06a}.
While \citet{burgasser07b} proposed a recipe for assigning spectral 
types to L subdwarfs by comparing with known solar-metallicity L dwarfs, 
no set of spectral standards is currently defined due to the few L subdwarfs 
known, so we consider our spectral type as tentative.

\subsection{Distance and tangential velocity}
\label{sdL:analysis_distance}

We have considered the absolute magnitude versus spectral type relations
given by \citet{cushing09} to estimate the distance of ULAS1350 as no
L5 subdwarf with known trigonometric parallax exist. Assuming a spectral
type of sdL5 and an uncertainty of one subtype, we derive M$_{J}$ = 12.715
mag (12.402--13.028 mag), yielding a mean distance of 111 pc with a possible 
range from 96 to 128 pc. Using the relations for $H$ and $K$, we find a
distance 25--30\% larger than for the $J$-band relation.
The uncertainty on the distance due to the relations
of \citet{cushing09} is 13--14 pc, implying that the uncertainty on the
distance is dominated by the error on the spectral classification.
To account for both uncertainties, we added the errors in quadrature, 
leading to a spectroscopic distance of 140$^{+30}_{-46}$ pc for ULAS1350\@.

Combining the proper motion (0.28 arcsec yr$^{-1}$) and the distance (140 pc) 
derived above, we derive a tangential velocity of 
186$^{+40}_{-61}$ km.s$^{-1}$ where the error bars account for the distance 
uncertainty. On the one hand, these values are $\sim$2--4 times larger 
than the mean values of tangential velocity reported for L dwarfs in the 
solar neighborhood \citep{vrba04,faherty09}. On the other hand, the 
transverse velocity of ULAS1350 is quite similar to the tangential velocities 
of the ultracool subdwarfs shown in Table 2 by \citet{schilbach09}. This 
indicates that ULAS1350 likely exhibits halo kinematics. As inferred from 
Figure \ref{fig_sdL:ccd}, state-of-the-art models predict a metallicity 
between [M/H]\,=\,$-$0.5 and $-$1.0 dex for ULAS1350 {although current
models do not reproduce accurately the color trends of field dwarfs 
mainly because condensate clouds are not included, resulting in offsets
between predicted and observed colors 
(Figure \ref{fig_sdL:ccd}; \citet{burgasser09a}).

%
%
\section{Discussion}
\label{sdL:discussion}
%


ULAS1350 was discovered in a 234 square area common to the SDSS and UKIDSS
surveys. Our search was limited to $J$ magnitudes brighter than 18.5\@.
Therefore, we can provide an estimate of the density of L5 subdwarfs in
the surveyed area. At the distance of 140 pc, the explored volume is
approximately 6600 pc$^{-3}$, implying a rough space density of
1.5$\times$10$^{-4}$ sdL5s per cubic parsecs. The density could, 
however, be larger as we have six other photometric candidates identified in 
the same area and we have not yet explored the faint end of the LAS\@. 
This discovery could also be fortuitous, yielding an overestimate
of the space density. Moreover, ULAS1350 could be a multiple system leading 
to the sampling of a larger volume. The inferred density is $\sim$20 times 
lower than the space density of solar-type field L dwarfs \citep[lower limit 
of 3.8$\times$10$^{-3}$ pc$^{-3}$;][]{cruz07}. It is also 10 times higher
than the density of M subdwarfs for M$_{V}$ = 5--10
\citep[1--3$\times$10$^{-5}$ pc$^{-3}$;][]{digby03}; we note however 
that our estimation is obtained for fainter and cooler objects, which may 
have smaller masses. This might indicate that the subdwarf mass function 
(in a linear scale) is flat or slowly rising toward the Hydrogen 
burning-mass limit. We remark that this is a tentative suggestion since it 
is based on the finding of one L5 subdwarf for which distance is not known 
precisely.
If we extrapolate our findings to the final area imaged by the LAS with SDSS 
photometry ($\sim$3000 deg$^{2}$), we would expect roughly 13
mid-L subdwarfs. This tentative number shows that our search may open new 
prospects of designing an accurate spectral classification in the L
dwarf regime including metallicity 
as a variable parameter \citep{kirkpatrick05}. With this aim, we have
conducted a cross-match of UKIDSS DR6 and SDSS DR7 following a Virtual 
Observatory methodology whose results will be presented in a
forthcoming paper.

%
%
\acknowledgments

N.L.\ was funded by the Ram\'on y Cajal fellowship number 08-303-01-02\@.
This work was partially funded by Spanish Ministry of Science grant number
AYA2007-67458 and the Consolider-Ingenio 2010 Program grant CSD2006-00070: 
``First Science with the GTC (http://www.iac.es/consolider-ingenio-gtc)''. 
NL thanks numerous discussions on subdwarfs with Richard Jameson during his 
post-doc in Leicester. We thank the GTC staff for their help as well as 
Adam Burgasser and Mike Cushing for providing the optical spectra of the 
known L subdwarfs used in this Letter. We thank the anonymous referee
for his/her prompt and useful report.

Based on observations made with the Gran Telescopio Canarias (GTC),
installed in the Spanish Observatorio del Roque de los Muchachos of the
Instituto de Astrofísica de Canarias, in the island of La Palma.

The United Kingdom Infrared Telescope is operated by the Joint
Astronomy Centre on behalf of the U.K.\ Science Technology and
Facility Council.

The SDSS is managed by the Astrophysical Research Consortium
for the Participating Institutions. The Participating Institutions
are the American Museum of Natural History, Astrophysical Institute
Potsdam, University of Basel, University of Cambridge, Case Western
Reserve University, University of Chicago, Drexel University, Fermilab,
the Institute for Advanced Study, the Japan Participation Group, Johns
Hopkins University, the Joint Institute for Nuclear Astrophysics, the
Kavli Institute for Particle Astrophysics and Cosmology, the Korean
Scientist Group, the Chinese Academy of Sciences (LAMOST), Los Alamos
National Laboratory, the Max-Planck-Institute for Astronomy (MPIA),
the Max-Planck-Institute for Astrophysics (MPA), New Mexico State
University, Ohio State University, University of Pittsburgh, University
of Portsmouth, Princeton University, the United States Naval Observatory,
and the University of Washington.

%
%
\begin{deluxetable}{l c c c c}
\tabletypesize{\footnotesize}
\setlength\tabcolsep{20pt}
\tablecolumns{5}
\tablecaption{Photometry and 
spectral type of ULAS J135058.86$+$081506.8 along with two 
comparison L subdwarfs and a field L dwarf.
}
\tablehead{
\colhead{} & \colhead{ULAS1350$^{a}$} & \colhead{SDSS1256$^{b}$} & \colhead{2MASS1626$^{b}$} & \colhead{SDSS1446$^{a,c}$}
}
\startdata
R.A.   &    13:50:58.86   &    12:56:37.10   &    16:26:20.34   &    14:46:00.60   \\
Decl.  & $+$08:15:06.8    & $-$02:24:52.5    & $+$39:25:19.0    & $+$00:24:52.0    \\
$i-z$  & 1.74$\pm$0.11    &    1.70$\pm$0.02 &    1.74$\pm$0.01 &    2.27$\pm$0.06 \\
$z-J$  & 1.55$\pm$0.07    &    1.61$\pm$0.02 &    1.71$\pm$0.01 &    2.92$\pm$0.03 \\
$J-K$  & $-$0.02$\pm$0.16 & $-$0.10$\pm$0.02 & $-$0.03$\pm$0.01 &    1.66$\pm$0.01 \\
$r$    & 24.30$\pm$0.66   &   21.80$\pm$0.11 &   20.61$\pm$0.03 &   23.19$\pm$0.24 \\
$i$    & 21.22$\pm$0.09   &   19.39$\pm$0.02 &   17.89$\pm$0.01 &   20.77$\pm$0.05 \\
$z$    & 19.48$\pm$0.06   &   17.68$\pm$0.02 &   16.15$\pm$0.01 &   18.50$\pm$0.03 \\
$Y$    & 18.60$\pm$0.05   &       ---        &      ---         &   16.90$\pm$0.01 \\
$J$    & 17.94$\pm$0.04   &   16.16$\pm$0.01 &   14.43$\pm$0.01 &   15.58$\pm$0.01 \\
$H$    & 18.08$\pm$0.10   &   16.06$\pm$0.01 &   14.46$\pm$0.01 &   14.66$\pm$0.01 \\
$K$    & 17.98$\pm$0.15   &   16.06$\pm$0.02 &   14.46$\pm$0.01 &   13.92$\pm$0.01 \\
SpT    & sdL5$\pm$1.0     &   sdL3.5$\pm$0.5 &   sdL4$\pm$0.5   &   dL5$\pm$0.5    \\
\enddata
 \label{tab_sdL:tab_phot}
\tablenotetext{a}{Optical photometry from SDSS DR7 (AB mag) and near-infrared photometry from UKIDSS LAS DR6 (Vega system)}
\tablenotetext{b}{Optical photometry from SDSS DR7 (AB mag) and near-infrared photometry from \citet{schilbach09}}
\tablenotetext{c}{Original discovery from \citet{geballe02}}
\end{deluxetable}

%
%
%
\begin{figure}
\includegraphics[width=\linewidth, angle=0]{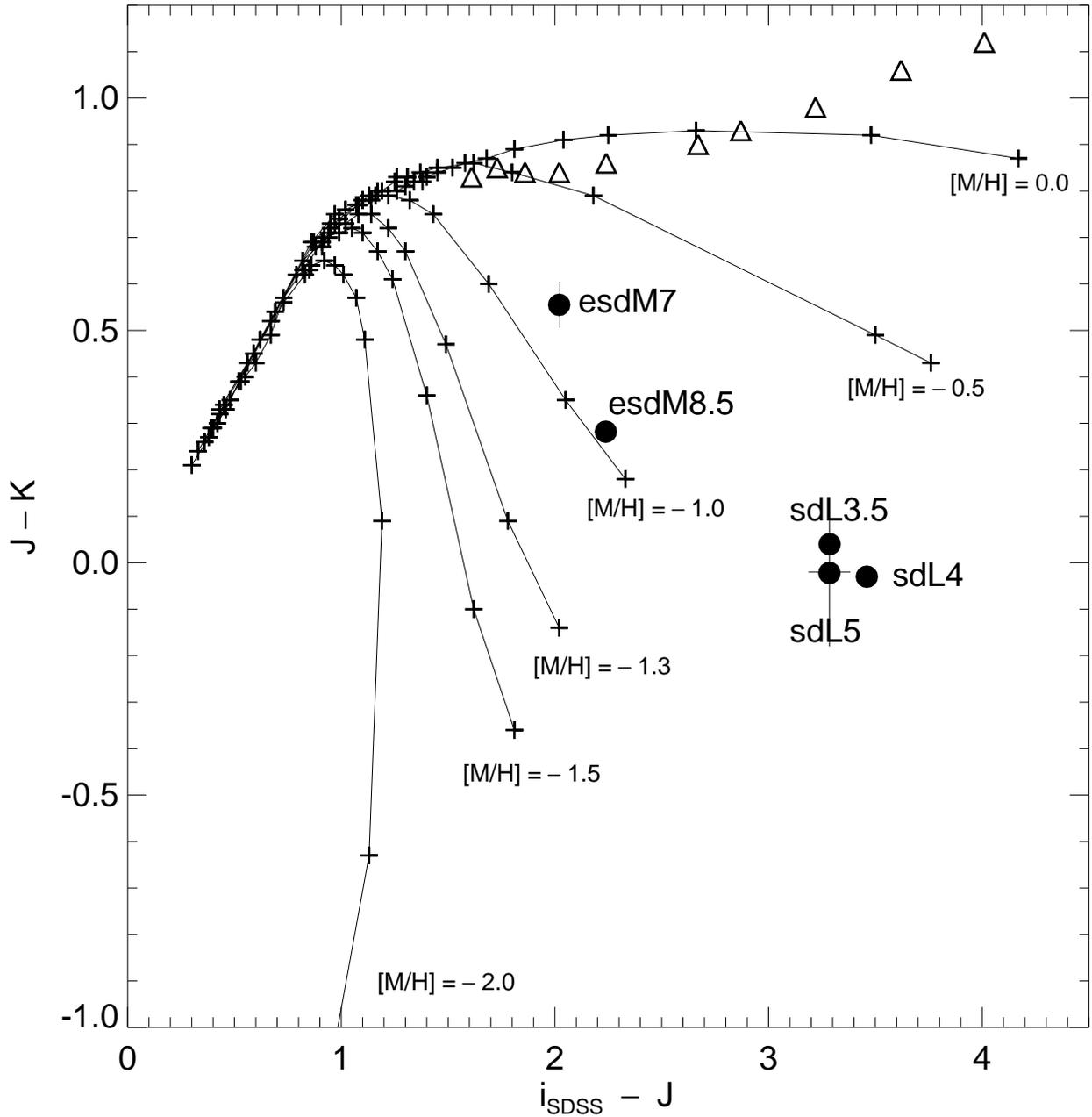}
   \caption{\scriptsize ($i-J$,$J-K$) diagram showing the location of two 
ultracool subdwarfs, SDSS J023557.61$+$010800.5 (esdM7) and 
SDSS J020533.75$+$123824.0 
(esdM8.5) from \citet{lepine08b} and three L subdwarfs as large dots: 
ULAS1350 (sdL5), 2MASS1626 (sdL4), and SDSS1256 (sdL3.5) with SDSS
photometry. Photometric error bars have been added for ULAS1350 and 
SDSS J023557.61$+$010800.5; the others being smaller than the size of the
symbols. Theoretical 10 Gyr tracks for different metallicities
are also shown \citep{baraffe97}. The colors of M0--M9 dwarfs
are overplotted as open triangles \citep{west08}; L dwarfs lie outside
this diagram because of their red $J-K$ colors. A similar
version of this plot was used by \citet{scholz04b} and \citet{burgasser09a}.
}
  \label{fig_sdL:ccd}
\end{figure}
%

%
%
\begin{figure}
 \begin{center}
  \includegraphics[width=\linewidth, angle=0]{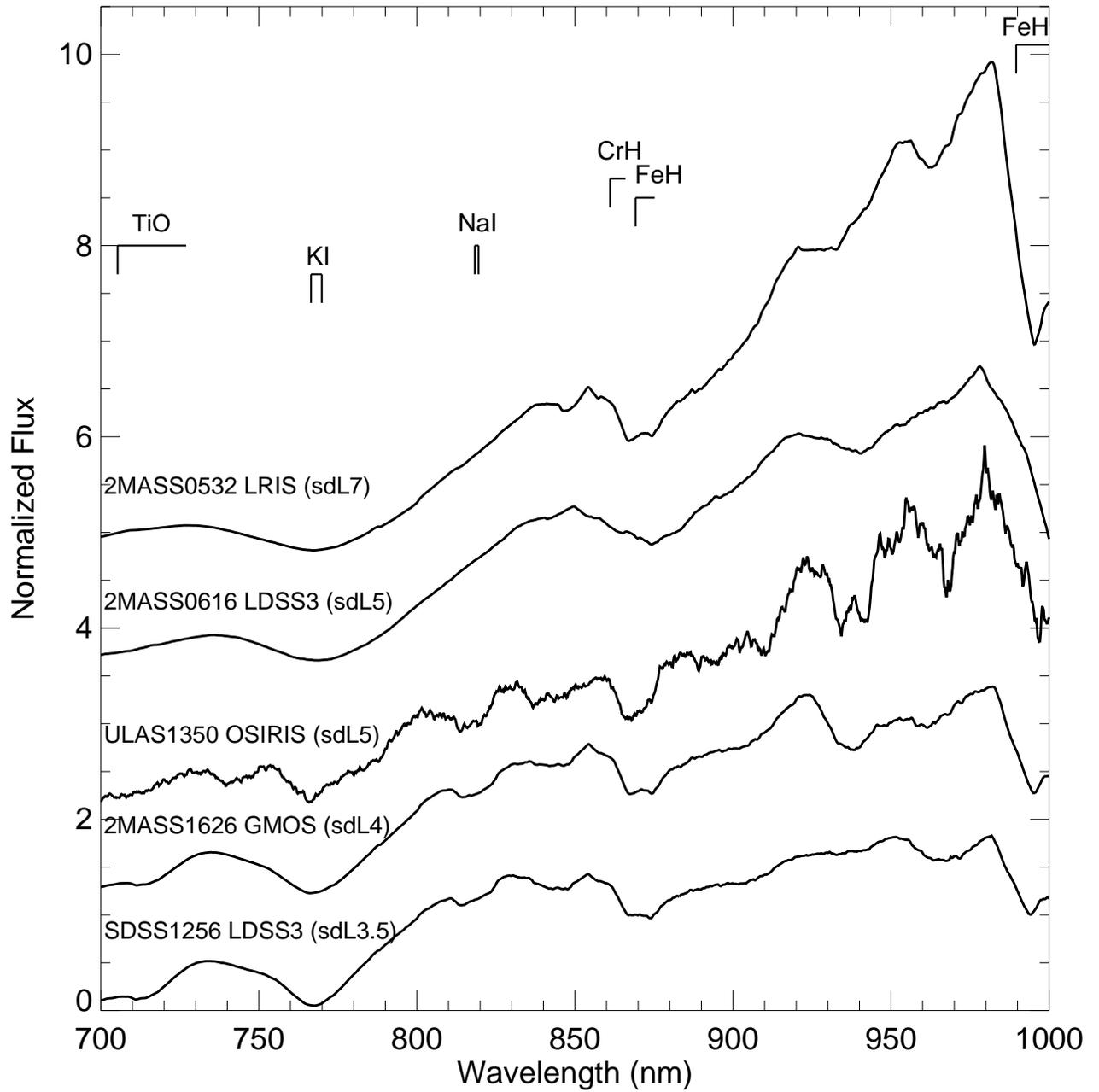}
  \caption{\scriptsize Smoothed ($R$ $\sim$ 10) optical spectrum of ULAS1350 
from the GTC\@. The signal-to-noise ratio of the spectrum
after smoothing is of the order of 20 over the 920--925 nm wavelength
range. Overplotted are the optical spectra of the four known 
L subdwarfs: 2MASS0532 \citep[Keck/LRIS; sdL7;][]{burgasser03b},
2MASS0616 \citep[Gemini/GMOS; sdL5;][]{cushing09},
2MASS1626 \citep[Gemini/GMOS; sdL4;][]{burgasser07b}, and
SDSS1256 \citep[Magellan/LDSS3; sdL3.5;][]{burgasser09a} 
degraded in wavelength resolution and smoothed in the same manner as
ULAS1350\@. Spectra have been normalized at 8200\AA{} and are shifted 
along the y-axis by 1.2 for clarity.
}
  \label{fig_sdL:sdL_spectrum}
 \end{center}
\end{figure}
%

%
%
\begin{figure}
 \begin{center}
  \includegraphics[width=\linewidth, angle=0]{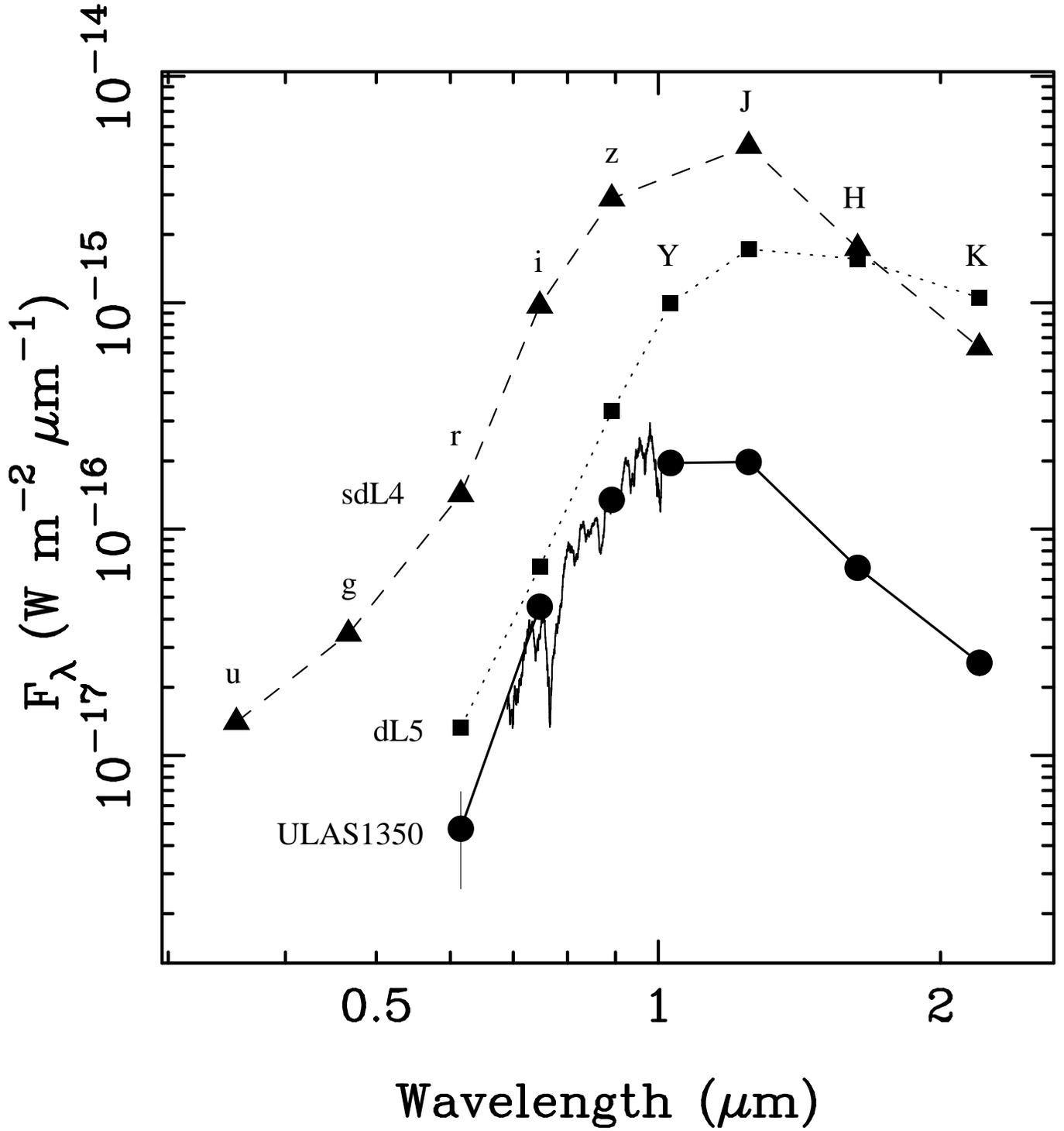}
  \caption{\scriptsize Observed SED of 
ULAS1350 formed by combining SDSS/UKIDSS photometry and the GTC spectrum 
(circles and solid line). For comparison purposes, the observed SEDs of the 
metal-depleted dwarf 2MASS1626 (sdL4; triangles and dashed line; optical
and near-infrared photometry from SDSS and \citet{schilbach09}, respectively) 
and the solar metallicity dwarf SDSS1446 (dL5; squares and dotted line;
photometry from SDSS and UKIDSS) are also shown. While the slope 
in the visible wavelengths are rather similar for solar and metal-depleted 
mid-L dwarfs, the SED appears quite different in the near-infrared. 
}
  \label{fig_sdL:sdL_SED}
 \end{center}
\end{figure}

\end{document}